\documentclass[]{spie}  

\usepackage{amsmath,amsfonts,amssymb}
\usepackage{graphicx}
\usepackage[colorlinks=true, allcolors=blue]{hyperref}

\title{INGOT Wavefront Sensor: from the optical design to a preliminary laboratory test}

\author[a,b,c]{Simone Di Filippo}
\author[a,b]{Davide Greggio}
\author[a,b]{Maria Bergomi}
\author[a,b]{Kalyan Radhakrishnan}
\author[a,b]{Elisa Portaluri}
\author[a,b]{Valentina Viotto}
\author[a,b]{Carmelo Arcidiacono}
\author[a,b]{Demetrio Magrin}
\author[a,b]{Luca Marafatto}
\author[a,b]{Marco Dima}
\author[a,b]{Roberto Ragazzoni}
\author[d,e]{Pierre Janin-Potiron}
\author[f]{Lauren Schatz}
\author[e]{Benoit Neichel}
\author[e]{Olivier Fauvarque}
\author[d,e]{Thirry Fusco}
\affil[a]{INAF - Osservatorio Astronomico di Padova, Vicolo dell’ Osservatorio 5, 35122, Padova, Italy}
\affil[b]{ADONI - ADaptive Optics National laboratory in Italy}
\affil[c]{Universit\'a  degli Studi di Padova - Dipartimento di Fisica e Astronomia, Vicolo dell’
Osservatorio 3, 35122, Padova, Italy}
\affil[d]{ONERA The French Aerospace Laboratory, F-92322 Ch\^atillon, France}
\affil[e]{Aix Marseille Univ, CNRS, CNES, LAM, Marseille, France}
\affil[f]{The University of Arizona, College of Optical Sciences, 1630 E University Blvd, Tucson, AZ 85719, USA}

\authorinfo{Further author information: E-mail: simone.difilippo@inaf.it}

\begin{document} 
\maketitle
\begin{abstract}
The Ingot wavefront sensor is a novel pupil-plane wavefront sensor, specifically designed to cope with the elongation typical of the extended nature of the Laser Guide Star (LGS). In the framework of the ELT, we propose an optical solution suitable for a Laser launch telescope, located outside the telescope pupil. In this paper, we present the current optical design, based on a reflective roof-shaped prism, which, at the level of the focal plane, splits the light from an LGS producing three beams. The three images of the telescope pupils can be then used for the retrieval of the first derivative of the wavefront. The 3D nature of such a device requires new alignment techniques to be determined theoretically and verified in the real world. A possible fully automated procedure, relying solely on the illumination observed at the three pupils, to align the prism to the image of the LGS is discussed. Careful attention needs to be put both on the telecentricity of the system and on the reference systems of the Ingot adjustments in the 3D space. This is crucial in order to disentangle all the possible misalignment effects. In this context, we devised a test-bench able to reproduce, in a scaled manner, the 3D illumination that the Ingot will face at the ELT, in order to validate the design and to perform preliminary tests of phase retrieval.
\end{abstract}


\section{INTRODUCTION}
The Ingot wavefront sensor (I-WFS), has been proposed in the ELT MAORY framework, as alternative to the classical concept of wavefront sensor, to cope with the typical elongation of the Laser Guide Star (LGS)\cite{Ragazzoni_2017}. This effect, is due to the intrinsic nature of the Sodium layer, which is located at about 90 km of altitude in atmosphere and extended for approximately 10 km. This layer, not only has a certain thickness, but also a certain density distribution, both varying in time. These properties, lead the LGS to be considered as an extended elongated object, with a given orientation in the sky. Considering a focal plane wavefront sensor, like the Shack-Hartman Wavefront Sensor (SH-WFS), the spot of the source will be only partially in focus, since the source extension along the optical axis is not negligible with respect to the LGS altitude. I-WFS aims to optimize the 1:1 matching between the source and the WFS used to sense it. The heart of the concept is to make the source image best focused on the element used to discriminate the parts of the wavefront affected by different local first-derivative, in order to increase the signal-to-noise ratio of the measurement. In the following, we discuss the current optical design of the I-WFS, and two different laboratory tests.
\section{ingot optical design}
The current I-WFS optical design is based on a 3 faces prism roof. In this configuration the LGS image is divided in three parts, instead of six as presented in the previous works\cite{deal,ext}. This new configuration, is called Ingot-3 WFS and, as before, the three separated beams are re-imaged into three corresponding pupils.

The Ingot-3 prism should take into account the following characteristics:
\begin{itemize}
    \item{Match the geometry of the LGS image and split the light at the level of the LGS focal plane into three beams that produce three images of the ELT entrance pupil}
    \item{Efficient light throughput}
    \item{Avoid overlapping of the output pupils}
\end{itemize}
According to the ELT and Sodium layer mean parameters, the LGS image tilt angle with respect to a focal plane perpendicular to the optical axis is 83.67$^{\circ}$. This angle is constant all over the field of view, being the input focal plane telecentric\cite{maory}. Instead, considering the F/5 input cone angle, the range of angles of incidence is (77.96-89.38)$^{\circ}$. With such a grazing incidence, a refractive solution results inefficient due to the high reflectivity at air-glass interface. Hence, the best option is to use a reflective prism.

\begin{figure}[h!]
    \centering
    {{\includegraphics[width=15cm]{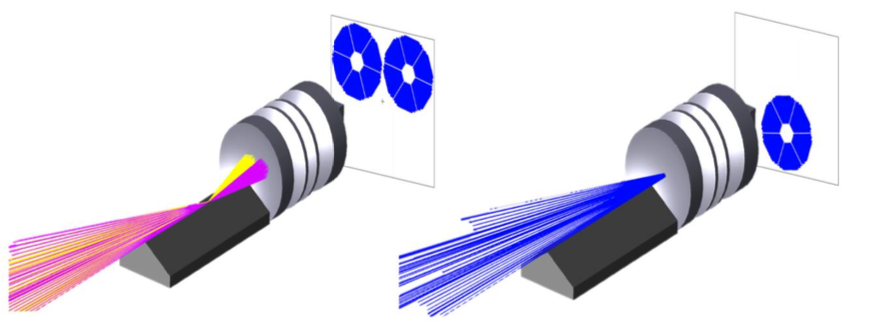}}}
    \caption{Conceptual layout of the ingot prism. Part of the LGS is focused by the reflecting ingot roof and forms two pupils while the remaining part of the LGS is focused after the ingot and is transmitted directly to the pupil re-imaging optics forming the third pupil}
    \label{fig:turbulence}
\end{figure}

\begin{figure}[h!]
    \centering
    {{\includegraphics[width=14cm]{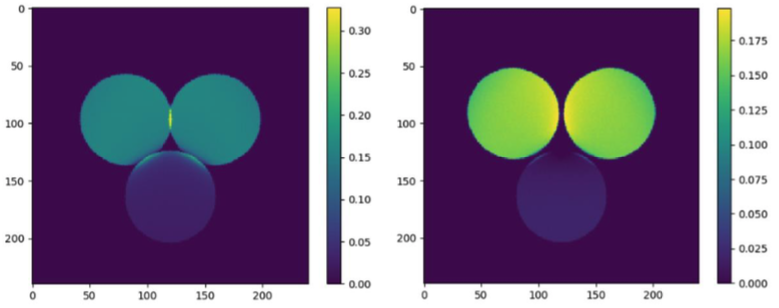}}}
    \caption{Image of the three pupils obtained by ray-tracing through the Ingot prism. Left: Ingot prism is placed on the focal plane of the LGS. Right: Ingot prism has been tilted by 0.5◦ with respect to the perfect focal plane of the LGS to separate the pupil images.}
    \label{fig:pupils}
\end{figure}

The solution adopted for the Ingot-3 prism is thus a reflective roof with its edge positioned along the axis of the LGS image. The roof is placed on the LGS image such to intercept only a portion of the LGS (which is reflected in two pupils) while the remaining portion of the LGS is not intercepted by the prism and will produce a third pupil. The 3 pupil images are then formed by a single objective placed after the ingot prism, as shown in Figure \ref{fig:turbulence}.
It is essential to define the apex angle of the prism in order to avoid overlapping of the output pupils. Hence, to achieve the proper separation value, the angle between the chief rays reflected by the two faces of the roof, should be greater than the input F/5 cone angle and the same should hold true for the angle between the transmitted and reflected chief rays. Unfortunately, with the E-ELT design parameters, the only possibility to avoid the pupil overlap is to offset the inclination of the prism with respect to the LGS image plane by a small angle. Tilting the prism by 0.5$^{\circ}$. from its nominal position, could be enough to remove the pupil overlap and have a clearance of about 1 sub-aperture between the re-imaged pupils, as shown in Figure \ref{fig:pupils}.
We stress here that this adjustment is only due to the characteristics of the E-ELT telescope (namely the distance of the laser launch telescope from the optical axis). Nevertheless, we do not expect any relevant impact of this small offset to the wavefront sensing, once the system is calibrated.
\section{ingot test bench optical design}
\begin{figure}[h!]
    \centering
    {{\includegraphics[width=17cm]{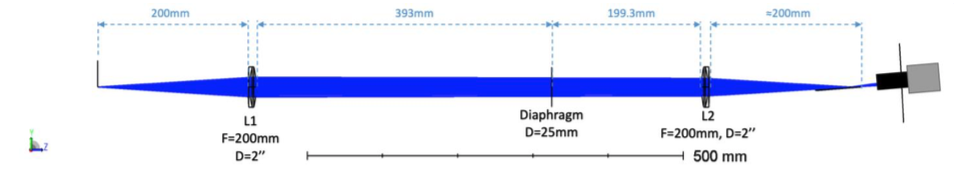}}}
    \caption{Optical Design of the I-WFS test bench.}
    \label{fig:design}
\end{figure}

\begin{figure}[h!]
    \centering
    {{\includegraphics[width=15cm]{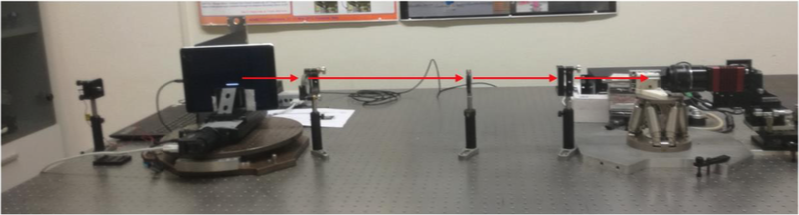}}}
    \caption{Ingot test bench.}
    \label{fig:banco}
\end{figure}

\begin{figure}[h!]
    \centering
    {{\includegraphics[width=15cm]{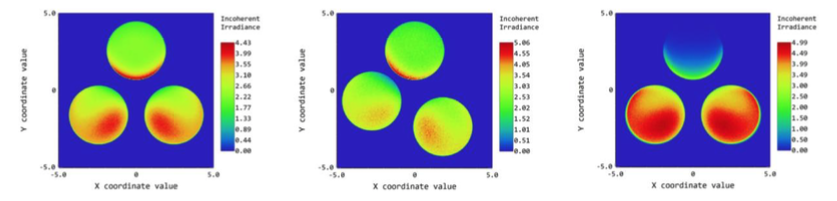}}}
    \caption{Ingot spot diagrams misalignment simulation. Left: Aligned configuration. Center: Tilt of 10◦ on X axis. Right: Shift of -25 mm on Z axis.}
    \label{fig:zemax}
\end{figure}

We designed a preliminary optical setup for an Ingot laboratory experiment in order to test the system and verify the alignment procedures and the signals recovered by the I-WFS in presence of aberrations. The optical configuration is shown in Figure \ref{fig:design}, and is composed by two lens with focal length $F_{1,2}$ = 200mm and by diaphragm with an aperture of D = 25mm. This configuration has been made in order to make the system telecentric in the object space. The system works at 1:1 magnification creating the image of the LGS source into the Ingot prism. The system has been designed taking into account the possibility to insert a transmitting phase screens in order to simulate turbulent layers.
In figure \ref{fig:banco} is shown the test bench. The LGS source, has been implemented using a monitor, on which is
reproduced as a white line, in order to simulate an elongated source. It’s stage is tilted by an angle of 4$^{\circ}$ with
respect to the optical axis with a length of $\simeq20-40mm$, which correspond to a Sodium layer thickness of $\simeq10-20km$. The Ingot, which is positioned on an Exapod stage, is tilted of angle $\alpha \simeq4.2^{\circ}$ with respect to the optical axis. 
In order to define an alignment procedure for the Ingot on the bench, it has been built a ray tracing simulator to explore all the possible misalignment cases, as shown in example in Figure \ref{fig:zemax}
\section{ingot test at loops bench at laboratoire d'astrophysique de marseille}
\begin{figure}[h!]
    \centering
    {{\includegraphics[width=11.5cm]{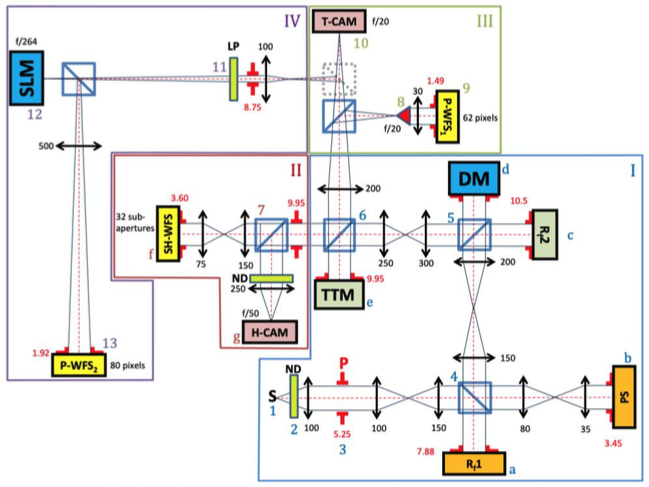}}}
    \caption{Schematic view of the LOOPS bench}
    \label{fig:LAM}
\end{figure}

The I-WFS has been simulated at the LOOPS Adaptive Optics test bench hosted at the Laboratoire d’ Astro- physique de Marseille (LAM)\cite{janin}, using a Spatial Light Modulator (SLM). Such device, placed in the focal plane, as shown in Figure \ref{fig:LAM}, is able to produce high definition phase masks that can mimic otherwise bulk device.

\begin{figure}[h!]
    \centering
    {{\includegraphics[width=16.5cm]{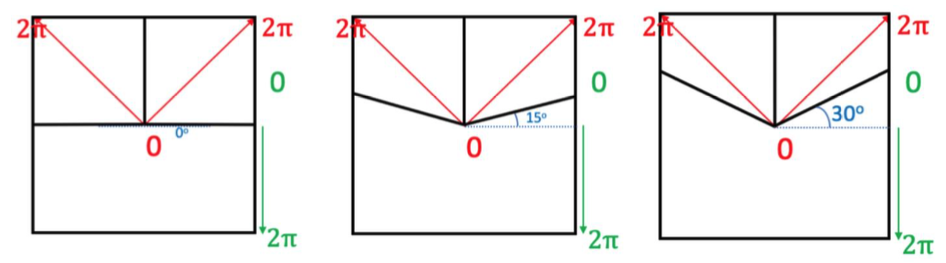}}}
    \caption{Ingot phase masks used to reproduce the three most representative cases. Left: Ingot seen from the closest sub-aperture to the laser launcher, corresponding to the source with aspect ratio of 1:1. Center: Ingot seen from a middle positioned sub-aperture, corresponding to the source with aspect ratio of 1:15. Right: Ingot seen from the farther sub-aperture with respect to the laser launcher, corresponding to the source with aspect ratio of 1:30.}
    \label{fig:masks}
\end{figure}

The LOOPS bench has made possible to test the I-WFS in quasi-real adaptive optics closed-loop operations. In order to reproduce the extended source, as the real LGS case, it has been used a fast tip-tilt mirror to introduce an elliptical modulation to the point source already available on the site. The length of the extended source, has been calculated in the ELT framework, considering three cases, from the less extended one (point source - with no elongation), to the more extended one (10km Sodium layer thickness), plus a case in the middle. These cases, characterized by an aspect ratio respectively of (1:1 - 1:15 - 1:30), correspond to the elongation observed by different telescope sub-apertures. Notice that also the I-WFS prism is seen under different projection, depending on the location of the sub-aperture. Hence, we have considered three different I-WFS phase masks mimicking the projection seen by the three sub-apertures. Ingot phase masks are shown in Figure \ref{fig:masks}

\begin{figure}[h!]
    \centering
    {{\includegraphics[width=10cm]{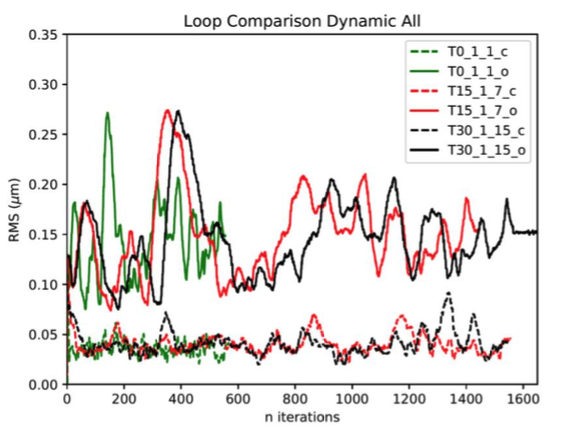}}}
    \caption{WFE measured from the SH camera for some combinations of Ingot masks and elongation. Solid lines are for open loop while dashed ones are for closed loop. The plots refer to the loops with dynamic disturbance, for Wavefront built from a combination of 65 Zernike modes}
    \label{fig:plot}
\end{figure}

In conclusion, for the different combinations of Ingot phase masks and modulated sources, it has been possible to collect both open and closed loop data, measuring the residual wavefront error from the SH camera. In Figure \ref{fig:plot}, are shown as a comparison, the open and closed loop data, for all the tested configurations when the dynamic disturbance produced from a combination of 65 Zernike modes is applied.
\bibliography{report} 

\begin{thebibliography}{1}

\bibitem{Ragazzoni_2017}
Ragazzoni, R., Portaluri, E., Viotto, V., Dima, M., Bergomi, M., Biondi, F.,
  Farinato, J., Carolo, E., Chinellato, S., Greggio, D., and et~al., ``Ingot
  laser guide stars wavefront sensing,'' {\em Proceedings of the Adaptive
  Optics for Extremely Large Telescopes 5}  (2017).

\bibitem{deal}
Viotto, V., Portaluri, E., Arcidiacono, C., Ragazzoni, R., Bergomi, M.,
  Filippo, S.~D., Dima, M., Farinato, J., Greggio, D., Magrin, D., and
  Marafatto, L., ``{Dealing with the cigar: preliminary performance estimation
  of an INGOT WFS},'' in [{\em Adaptive Optics Systems
  VI}{\nolinebreak\hspace{0.1em}]},  Close, L.~M., Schreiber, L., and Schmidt,
  D., eds.,  {\bf 10703},  276 -- 283, International Society for Optics and
  Photonics, SPIE (2018).

\bibitem{ext}
Ragazzoni, R., Greggio, D., Viotto, V., Filippo, S.~D., Dima, M., Farinato, J.,
  Bergomi, M., Portaluri, E., Magrin, D., Marafatto, L., Biondi, F., Carolo,
  E., Chinellato, S., Umbriaco, G., and Vassallo, D., ``{Extending the pyramid
  WFS to LGSs: the INGOT WFS},'' in [{\em Adaptive Optics Systems
  VI}{\nolinebreak\hspace{0.1em}]},  Close, L.~M., Schreiber, L., and Schmidt,
  D., eds.,  {\bf 10703},  1106 -- 1111, International Society for Optics and
  Photonics, SPIE (2018).

\bibitem{maory}
Ciliegi, P., Diolaiti, E., Abicca, R., Agapito, G., Aliverti, M., Arcidiacono,
  C., Auricchio, N., Balestra, A., Baruffolo, A., Bellazzini, M., Bonaglia, M.,
  Bregoli, G., Brissaud, O., Busoni, L., Carlotti, A., Cascone, E., Correia,
  J.-J., Cortecchia, F., Cosentino, G., D'Orazi, V., Dall'Ora, M., Caprio,
  V.~D., Rosa, A.~D., Delboulbé, A., Antonio, I.~D., Rico, G.~D., Dolci, M.,
  Esposito, S., Fantinel, D., Feautrier, P., Fiorentino, G., Foppiani, I.,
  Giro, E., Gluck, L., Grani, P., Greggio, D., Hénault, F., Jocou, L., Penna,
  P.~L., Lafrasse, S., Lauria, M., Coarer, E.~L., Louarn, M.~L., Lombini, M.,
  Magnard, Y., Magrin, D., Maiorano, E., Mannucci, F., Marchetti, E., Maurel,
  D., Michaud, L., Moraux, E., Morgante, G., Moulin, T., Oberti, S., Pariani,
  G., Patti, M., Plantet, C., Podio, L., Puglisi, A., Rabou, P., Ragazzoni, R.,
  Redaelli, E., Riva, M., Rochat, S., Roussel, F., Roux, A., Salasnich, B.,
  Saracco, P., Schreiber, L., Spavone, M., Stadler, E., Sztefek, M.-H.,
  Terenzi, L., Valentini, A., Ventura, N., Vérinaud, C., and Zaggia, S.,
  ``{MAORY for ELT: preliminary design overview},'' in [{\em Adaptive Optics
  Systems VI}{\nolinebreak\hspace{0.1em}]},  Close, L.~M., Schreiber, L., and
  Schmidt, D., eds.,  {\bf 10703},  336 -- 345, International Society for
  Optics and Photonics, SPIE (2018).

\bibitem{janin}
Janin-Potiron, P., Chambouleyron, V., Schatz, L., Fauvarque, O., Bond, C.~Z.,
  Abautret, Y., Muslimov, E.~R., Hadi, K.~E., Sauvage, J.-F., Dohlen, K.,
  Neichel, B., Correia, C.~M., and Fusco, T., ``{Adaptive optics with
  programmable Fourier-based wavefront sensors: a spatial light modulator
  approach to the LAM/ONERA on-sky pyramid sensor testbed},'' {\em Journal of
  Astronomical Telescopes, Instruments, and Systems}~{\bf 5}(3),  1 -- 10
  (2019).

\end{thebibliography}
\bibliographystyle{spiebib} 

\end{document}